\documentclass[preprint]{sig-alternate-05-2015}

\usepackage{xcolor}
\usepackage{mathtools}
\usepackage{amsmath}
\usepackage[vlined]{algorithm2e}
\usepackage{multirow}
\usepackage{subcaption}

\begin{document}

\toappear{}
\title{The Tersoff many-body potential: \\ Sustainable performance through vectorization}
\numberofauthors{3}
\author{
\alignauthor
Markus H\"ohnerbach\\
\affaddr{RWTH Aachen University}
\alignauthor
Ahmed E. Ismail\\
\affaddr{West Virginia University}
\alignauthor
Paolo Bientinesi\\
\affaddr{RWTH Aachen University}
}
\maketitle
\section{Introduction}

We extend the LAMMPS molecular dynamics program with a new, vectorized implementation of the Tersoff potential \cite{tersoff}.
Given the well-established and well-studied mechanisms for parallelism in molecular dynamics programs \cite{Plimp}, our efforts focus on vectorization as a further step to fully utilize available hardware.
This contribution describes how our implementation achieves sustainable performance across a number of architectures, most notably the Xeon Phi coprocessor, using vectorization.

On current architectures, vectorization contributes greatly to the system's peak performance;
this is true for CPUs with the SSE or AVX instruction set
extensions, and especially for machines with wide vectors, such as the Xeon Phi.
Many successful open-source molecular dynamics packages---e.g.~Gromacs, NAMD, LAMMPS and ls1 mardyn~\cite{user-intel, gromphi, ls1phi}---take advantage of vectorization.
Implementation methods vary between hand-written assembly, intrinsics (compiler-provided functions that closely map to machine instructions), and annotations that guide the compiler's optimization \cite{pragma}.

Typically, only the parts of the calculation that consume a large fraction of the total runtime are optimized;
among them, the neighbor list construction and the force calculation.
For most simulations, the forces are derived from pair potentials, such as the Coulomb or Lennard-Jones potentials.
Indeed, vectorized implementations of the force calculation due to pair potentials are found in many molecular dynamics programs.

However, some applications, especially in materials science, require many-body potential formulations.
With these, the force between two atoms does not depend solely on the distance between them, but also on the relative position of the surrounding atoms.
For many-body potentials, the force calculation is not vectorized in the available molecular dynamics programs.
There has been previous work on implementing many-body potentials on the GPU \cite{gpua, gpub, gpuc}, but not on more conventional architectures with vectorization support.

\section{The Tersoff potential}\label{sec:ters}

The Tersoff potential is an example of a many-body potential \cite{tersoff}.
It is an ideal target for a first vectorization attempt:
relatively simple structure, widely used, and still challenging to optimize.

The Tersoff potential is given by the following formulas (omitting trivial definitions):

\begin{align}
V & = \sum_i \sum_{j : r_{ij} < r_C} \overbrace{f_C(r_{ij}) \left[ f_R(r_{ij}) + b_{ij}f_A(r_{ij}) \right]}^{V(i, j, \zeta_{ij})},\label{form:ters-1}\\
b_{ij} & = \vphantom{\sum_{r_j}}(1 + \beta^\eta\zeta_{ij}^\eta)^{-\frac{1}{2\eta}}, \eta\in\mathbb{R}\\
\zeta_{ij} & = \vphantom{\sum_{r_j}}\sum_{k : r_{ik} < r_C} \underbrace{f_C(r_{ik}) g(\theta_{ijk}) \exp(\lambda_3 (r_{ij} - r_{ik}))}_{\zeta(i, j, k)}.\label{form:ters-3}
\end{align}
Formula~\ref{form:ters-1} indicates that two forces act between each pair of atoms $(i, j)$, an attractive force modeled by $f_A$, and a repulsive force modeled by $f_R$.
Both depend only on the distance $r_{ij}$ between atom $i$ and atom $j$.
The bond-order factor $b_{ij}$, however, depends on all the other atoms $k$ surrounding atom $i$, by means of their distance $r_{ik}$, and angles $\theta_{ijk}$.
Since the contribution of the $(i, j)$ pair depends on the other atoms $k$,
the Tersoff potential is a many-body potential.
Although the empirical nature of the potential means that many parameter
lookups are necessary, the functions within it ($f_R, f_A, f_C, g, \exp$) are
expensive to compute, which makes it
a good target for vectorization.

Algorithm~\ref{algo:ters-lammps} implements the force and energy calculation, derived from Equations~\ref{form:ters-1}--\ref{form:ters-3}, in terms of the functions $V(i, j, \zeta)$ and $\zeta(i, j, k)$.
In a molecular dynamics simulation, one needs to not only calculate the potential energy, but also the associated forces on the atoms.
Consequently, the algorithm calculates the derivatives of the potential with respect to the atom positions.
Algorithm~\ref{algo:ters-lammps} describes the implementation provided by LAMMPS:
For all $(i, j)$ pairs of atoms, first accumulate $\zeta_{ij}$, then update the forces based on the $V(i, j, \zeta_{ij})$ term, and finally perform the updates due to the $\zeta(i, j, k)$ terms.
This means that for each atom $i$ with a neighbor list of length $n$, the algorithm visits each neighbors $2n$ times (in total $2n^2$).

\begin{algorithm}
\caption{Tersoff potential and forces}\label{algo:ters-lammps}
\For{$i$ in local atoms of the current thread}
{
 \For{$j$ in atoms neighboring $i$}
 {
  $\zeta_{ij} \gets 0$\;
  \For{$k$ in atoms neighboring $i$}
  {
    $\phantom{F^{ij}_k}\mathllap{\zeta_{ij}} \gets \phantom{F^{ij}_k}\mathllap{\zeta_{ij}} +\, \zeta(i, j, k)$\;
  }
  $E \gets E + V(i, j, \zeta_{ij})$\;
  $F_i \gets F_i - \partial_{x_i} V(i, j, \zeta_{ij})$\;
  $F_j \gets F_j - \partial_{x_j} V(i, j, \zeta_{ij})$\;
  $\delta\zeta \gets \partial_{\zeta}V(i, j, \zeta_{ij})$\;
  \For{$k$ in atoms neighboring $i$}
  {
    $\phantom{F_k}\mathllap{F_i} \gets \phantom{F_k}\mathllap{F_i} - \delta\zeta\cdot\,\partial_{x_i}\zeta(i, j, k)$\;
    $\phantom{F_k}\mathllap{F_j} \gets \phantom{F_k}\mathllap{F_j} - \delta\zeta\cdot\,\partial_{x_j}\zeta(i, j, k)$\;
    $F_k \gets F_k - \delta\zeta\cdot\,\partial_{x_k}\zeta(i, j, k)$
  }
 }
}
\end{algorithm}

For the following discussions, it is important to keep the loop structure of Algorithm~\ref{algo:ters-lammps} in mind:
There is an outer loop $I$ over all atoms, then a loop over all neighbors $J$,
and inside the latter two more loops over the neighbors $K$.
The first optimization is to restructure the algorithm such that $\zeta$ and its derivatives are only computed once, in the first loop, and only the product with $\delta\zeta$ is performed in the second loop.
Since $\zeta$ and its derivatives---naturally---share terms, this modification has a measurable impact on performance.

The main vectorization challenge however is the extremely short neighbor list that is typical for many-body potentials:
In a representative simulation run, the neighbor list for any given atom will rarely contain more than four atoms.
However, to avoid frequent rebuilds of the neighbor list, this may include a number of additional atoms;
in order to not waste slots in the actual potential calculation on these skin
atoms, these have to be excluded efficiently.
The exact exclusion mechanism constrains the vectorization, because it
determines the amount of necessary control flow in the vectorized regions, and
the strategy to update the forces.

With most common pair potential formulations, one would be able to increase the neighbor list's length by extending the cutoff radius---the maximum distance for an atom to be included in another atom's neighbor list.
Increasing the cutoff shifts work from the long-ranged force calculation to the short-ranged calculation, and improves accuracy.
However, the potential presented here does not have a long-range component, and does not even have a cutoff in the conventional sense:
The cutoff instead is an intrinsic property of the chosen parametrization.

\section{Optimization}\label{sec:opt}

From the previous discussion, it is clear that the most straightforward approach---vectorizing the iteration through the neighbor list---will not yield desired speedups.
The reason is that most entries in the neighbor list are part of the skin region.
They do not actually contribute to the calculation.
Including them would lead to many calculations that would in the end be masked out.
The alternative is to examine all entries to determine if they are in the cutoff region, and to process only those entries batch-wise, using a vectorized method.
The packing addresses the ``sparsity'' of the neighbor list, as it exposes the short length of the resulting packed neighbor list. 
The missing component is the vectorized algorithm for batch-wise processing:
(1) vectorize along the $J$ loop, and (2) vectorize along both the $I$ and the $J$ loop.

In (1), the $J$ loop is vectorized, which is the ``middle'' loop; in the more common pair potential case it would be the innermost loop, and thus the natural candidate for vectorization.
The advantage of this approach is that the atom $i$ is constant across all lanes.
As the $K$ loops iterate through the neighbor list of atom $i$, the same neighbor list is traversed across all lanes, leading to an efficient vectorization.
However, with long vectors and short neighbor lists, this approach is destined to fail on the Xeon Phi.

Method (2) handles the previous shortcoming:
Vectorizing over both the $I$ and the $J$ loop, effectively considering atom pairs $(i, j)$, again allows for packing.
Given that the $I$ loop iterates over all the atoms in the simulation, this strategy additionally makes it possible to exploit arbitrarily large vectors.
In contrast to (2), with the vectorization of the $I$ loop, atom $i$ is not constant across all lanes.
Consequently, the innermost loops will iterate over the neighbor lists of different $i$, leading to a more involved iteration scheme.
Even if this iteration is very efficient, it can not attain the same performance of an iteration scheme where all vector lanes iterate over the same neighbor list.
The vectorization of the $I$ loop invalidated a number of assumptions of the algorithm:
$i$ and $k$ are always identical across all lanes, while $j$, coming from the same neighbor list, is always distinct.
Without these assumptions, special care has to be taken when accumulating the forces.
For the program to be correct under all circumstances, the updates have to be serialized.

Whether the disadvantages outweigh the advantages or not is primarily a question of amortization.
The answer to this question depends on the used floating point data type, the vector length, and the features of the underlying instruction set.

\section{Implementation}

Implementations of the two algorithms described in section~\ref{sec:opt} where added to the LAMMPS molecular dynamics program, building upon the USER-INTEL package \cite{user-intel}, which provides basic support for offloading and vectorized implementations.
In production runs, the Xeon Phi is accessed via offloading, in the same way any other accelerator is;
for benchmarking purposes, the program runs natively to isolate the effect of vectorization.
Parallelism is implemented according to the standard paradigm of the package:
MPI is used for a domain decomposition, and atom decomposition is used at the OpenMP level.

The Intel Compiler version 16.0 is used.
Initially, we set out to rely on source code annotations and compiler-supported optimization.
However, it seems impossible to create a compiler-vectorized implementation that follows our optimization strategies.
In particular, the compiler model assumes that all vector lanes write data to
distinct memory locations; 
in the innermost loop of our implementations, this assumption is not satisfied.
Instead, as a first prototype, we created a Xeon Phi double precision intrinsics version.
The usage of intrinsics is problematic, because it leads to verbose, hard to read and debug-unfriendly code.
Given that the creation of the prototype was a considerable effort, creating
further intrinsics programs to cover additional platforms is not sustainable.

For the final product, both support for different instruction sets and
floating point precision settings are necessary.
It is crucial to also support CPU instruction sets to balance the load between host and accelerator.
Additionally, such a code enables us to evaluate the influence of vector lengths and instruction set features on the code performance.
Looking at all combinations of instruction sets, data types and vectorization variants, it is infeasible to implement them in intrinsics.
Instead, a single, tested, correct algorithm is created, and paired with a tested, correct vectorization support library.
Consequently, instead of implementing the Tersoff potential's algorithm $n\cdot m$ times ($n$ architectures and $m$ precision modes), we only need to implement simple building blocks in the support library.

The vectorization support library is implemented using C++ templates which are specialized for each targeted architecture.
The library contains implementations for single, double and mixed precision using a variety of instruction set extensions:
Scalar, SSE4.2, AVX, AVX2, IMCI (the Xeon Phi Knights Corner instruction set), as well as experimental support for AVX-512 and Cilk array notation.

The array notation implementation can help debugging the code as it can simulate arbitrary vector lengths.
Adding a new architecture is straightforward, although tuning might take some time.
Tuning is simplified as the library is only a very thin abstraction layer, and provides a number of bigger building blocks such as wide gather-and-transpose operations which can be optimized in one go.

Contrary to most other vector libraries, which allow the programmer to pick a
vector length that may be emulated internally, this vector library allows for algorithms that are oblivious of the used vector length.

\section{Benchmarking}

The benchmarks use a simplified carbon nanotube stretching simulation as the model problem.
In a carbon nanotube, each atom has roughly three nearest neighbors.
Consequently, Section~\ref{sec:ters}'s and Section~\ref{sec:opt}'s considerations about small neighbor lists apply here.

Subsections~\ref{ssec:vecphi} and \ref{ssec:veccpu} evaluate the vectorization quality isolated from all other effects.
The quality is measured by running on the bare hardware with a single thread, and comparing the normal LAMMPS version against our scalar-optimized version and our vectorized-optimized version.
Measuring the runtime of all these versions allows us to isolate the performance gains due to vectorization from those due to other optimizations.
Running with a single thread allows us to eliminate threading and offloading overheads.
Continuing, Subsection~\ref{ssec:full} also compares entire nodes of varying system configuration and their respective computing power for this specific problem.
For this purpose, a longer, more complicated variant of the same simulation is run, utilizing each available machine at its fullest.

\subsection{Vectorization on the Xeon Phi coprocessor}\label{ssec:vecphi}

For the Xeon Phi coprocessor, only the ``I'' algorithm is considered.
The ``J'' algorithm is not competitive due to its shortcomings, that were discussed in the previous section.
The benchmark runs natively on a single core of a mid-range 5110P Xeon Phi coprocessor, and both single and double precision performance is measured.
All comparisons are relative to the LAMMPS baseline, which only provides a double precision implementation, limiting the validity of single precision speedups.

Table~\ref{stab:vphi-s} indicates that scalar optimizations lead to a 1.5x speedup in double precision, and that vectorization is responsible for another factor of four.
As the Phi's vector unit is eight double elements wide, this means that the code achieved a vectorization efficiency---speedup per vector length---of 50\%.

For single precisions, all the speedups are higher than for double precision, while the vector efficiency decreases to 33\%.
The reason for this drop is that certain operations have to take place on a per-lane basis.
Unfortunately there is no single precision LAMMPS version to compare against.

\begin{table}
\centering
\begin{subtable}{.5\textwidth}
\caption{Timings (in seconds)}
\centering
\begin{tabular}{l|rrr}
Precision & LAMMPS & I-Scalar & I-Vec\\
\hline
double & 88.72 & 58.04 & 14.18\\
single &---     & 45.59 & 8.56
\end{tabular}
\end{subtable}
\begin{subtable}{.5\textwidth}
\centering
\caption{Speedup \& Vector Efficiency}\label{stab:vphi-s}
\begin{tabular}{l|rrrr}
Precision & $\frac{\textsf{LAMMPS}}{\textsf{I-Scalar}}$ & $\frac{\textsf{LAMMPS}}{\textsf{I-Vec}}$ & $\frac{\textsf{I-Scalar}}{\textsf{I-Vec}}$ & $\frac{\textsf{I-Scalar}}{\textsf{I-Vec}\cdot\textsf{VL}}$\\[0.1em]
\hline
double & 1.53 & 6.26 & 4.09 & 0.51\\
single & (1.95) & (10.36) & 5.32 & 0.33
\end{tabular}
\end{subtable}
\caption{Vectorization experiment on a Xeon Phi 5110P, using the ``I'' algorithm running directly on the coprocessor without threading.}
\end{table}

\subsection{Vectorization on the Xeon processor}\label{ssec:veccpu}

\begin{table}
\centering
Timings (in seconds) \& Speedups
\\[0.1em]
\begin{tabular}{l|rr|rr}
Arch.& ``I'' & ``J''& $\frac{\textsf{LAMMPS}}{\textsf{``I''}}$ & $\frac{\textsf{LAMMPS}}{\textsf{``J''}}$ \\[0.1em]
\hline
LAMMPS & \multicolumn{2}{c}{28.23} & \multicolumn{2}{|c}{1}\\
Scalar& 18.63 & 14.7 & 1.52 & 1.91\\
SSE  &  37.15 & 21.3 & 0.76 & 1.32\\
AVX  &  23.92 & 12.5 & 1.18 & 2.25\\
AVX2 &  16.59 & 10.9 & 1.70 & 2.59
\end{tabular}
\caption{Vectorization experiment on a Xeon E5-2680 v3 processor (Haswell generation), in double precision without threading}\label{tbl:xeon}
\end{table}

Our second benchmark is performed on a Xeon processor from the Haswell generation.
For the evaluate of vectorization, the code is run on a single core.
On this particular processor, the benchmark can run with 128-bit vectors and 256-bit vectors, and also utilize AVX2 instructions.
Given this multitude of options, only double precision measurements are given.
Consequently, the code has to deal with vector lengths of 2 (SSE) and 4 (AVX/AVX2) respectively.

Table~\ref{tbl:xeon} contrasts the runtime of both the ``I'' algorithm and the ``J'' algorithm, as well as the associated speedups.
Contrary to the Xeon Phi, the majority of speedups here comes from our sequential optimizations.
The speedup from vectorization is lower than expected on all of the targeted instruction sets.

For SSE, no speedup is achieved at all in double precision relative to the scalar version.
This is surprising because the vectorized code performs its calculation using the same instructions as its scalar counterpart.
There are a number of reasons why the scalar code outperforms the vectorized one, including the lack of efficient masking and the need to spend time in gather operations.
To read the correct parameters for an atom interaction, the algorithm has to perform a masked gather operation, which only has hardware support with the beginning of AVX2.
Otherwise, it has to rely on an expensive emulation.

The same argument also applies to AVX, however the longer vectors amortize some of that cost.
AVX actually lacks a number of operations in comparison with SSE, such as integer addition and multiplication, that need to be emulated.

AVX2 relieves most of these problems, as it provides integer operations and gather instructions.
As such, performance should be similar to the Xeon Phi, with a vector efficiency of 50\%, i.e. a speedup of 2 against the scalar optimized code.
That speedup however is just 1.35.
Reasons might include the lack of dedicated mask registers or the comparably slow gather operation (as opposed to the Phi).

In the future, gather operations could be reduced by explicitly specializing the kernel to the number of atoms types:
For up to four/eight types, gather instructions could be replaced with cheaper permutation operations.

\subsection{Full system comparison}\label{ssec:full}

\begin{table}
\centering
Timings (in seconds)
\\[0.1em]
\begin{tabular}{ll|rr}
System & & double & single\\
\hline
\multirow{2}{*}{Sandy Bridge} & LAMMPS & 395.89 & \\
                              & Vec    & 250.02 & 229.65\\
\hline
Phi                           & Vec.   & 170.88 & 125.14\\
\hline
\multirow{2}{*}{Haswell}      & LAMMPS & 182.43 & \\
                              & Vec.   & 136.99 & 103.16\\
\end{tabular}
\caption{Full system comparison: 
Measured runtimes}\label{tbl:full}
\end{table}

Finally, this section provides a comparison of complete systems:
These are one Haswell system, with two E5-2680 v3 (24 cores in total), one Sandy Bridge system, with two E5-2450 (16 cores in total), and the Xeon Phi 5110P (60 cores in total), connected to the Sandy Bridge system.
As opposed to the previous sections, the benchmark utilizes all cores of the respective system.
The production version of the code is used, which means that the Xeon Phi coprocessor is accessed via offloading.
To keep force reduction times due to atom decomposition low, a mixture of MPI ranks and threads on the Xeon Phi per rank are chosen.
The optimal parameters in this case are 8 MPI ranks, with 29 threads each.
While part of the domain decomposition work from the MPI parallelization is done on the host, the force and energy calculation is performed exclusively on the Xeon Phi.
As opposed to the CPUs, the Xeon Phi benefits from oversubscription: We run approximately four threads per core.

All the following measurements are ``production'' runs that the full systems at their best vectorization setting, with parallelization and realistic experimental conditions.
For the Xeon Phi, this includes the overhead due to offloading.
The baseline for the following comparisons is LAMMPS' OpenMP support.
A direct comparison is not possible for the Xeon Phi, as the package does not support offloading.

As expected from the previous sections, the resulting measurements in Table~\ref{tbl:full} show that our optimizations deliver a sizable performance improvement against the comparable LAMMPS variant on any system.
The aim of the benchmark is to incorporate all the overheads present in a real-world simulation.
Both parallelism and offloading introduce overheads, that were not present in the previous measurements.
As such, the speedups are expected to be lower than the speedups achieved using vectorization alone.

The ordering among the systems with respect to runtime is not unexpected given their release dates and computational power.
The Sandy Bridge system is slowest, followed by the Xeon Phi, and the fastest system is the two years younger Haswell system.
While peak floating point performance is an imperfect measure,  it offers some explanation of the ordering.
The theoretical peak FLOPS (DP) of the systems are 269 GFLOPS (DP) for Sandy Bridge, 1011 GFLOPS (DP) for the Xeon Phi, and 960 GFLOPS (DP) for Haswell.
Although our code does run on the prototype of the ``Knights Landing'' chip, hard performance data is not available yet.
Given that the chip is expected to deliver around 3000 GFLOPS (DP), it seems reasonable that it will outperform the Haswell system.

\section{Conclusion}

We showed that vectorization can achieve considerable speedups also in
complicated simulations which do not immediately lend themselves to the SIMD paradigm.
Indeed, even a many-body potential with a very short neighbor list benefits from vectorization, especially when using accelerator hardware such as the Xeon Phi coprocessor.
To implement the optimized algorithm in a portable way, the abstraction
library has proved useful, because it allows a clean division between the
algorithm implementation and the hardware support for vectorization.
The ideas behind our optimizations were described, and their effectiveness was
validated by means of realistic use cases.
Our results suggest that the upcoming generation of Xeon Phi chips will lead
to another increase in performance.

\section{Acknowledgments}

The authors gratefully acknowledge financial support from the Deutsche Forschungsgemeinschaft (German Research Association) through grant GSC 111, and from Intel via the Intel Parallel Computing Center initiative.
We would like to thank Marcus Schmidt for providing the benchmark used in this work.

\section{Open Source}

The associated code is available at \cite{gh}.

\end{document}